\documentclass{article}
\usepackage[preprint]{spconf}
\usepackage{amsmath,graphicx}
\usepackage{booktabs}
\usepackage{graphicx}
\usepackage{caption}
\usepackage{subcaption}
\usepackage{lipsum}
\usepackage{amssymb}
\usepackage{amsthm}
\usepackage{fancyhdr}

\newcounter{mycounter}  
\newenvironment{noindlist}
 {\begin{list}{\arabic{mycounter}.~~}{\usecounter{mycounter} \leftmargin=0em \itemindent=0em \labelwidth=0em \labelsep=0em}} 
 {\end{list}}

\copyrightnotice{\copyright\ IEEE 2022}
\toappear{To appear in {\it Proc.\ ICASSP2022, May 22-27, 2022, Singapore}}

 \title{Multi-View Self-Attention Based Transformer for Speaker Recognition}
\name{Rui Wang$^{1,\dag}$,  Junyi Ao$^{2,3,\dag}$,  Long Zhou$^4$, Shujie Liu$^4$, Zhihua Wei$^1$, Tom Ko$^2$, Qing Li$^3$, Yu Zhang$^2$\thanks{$^{\dag}$Equal contribution. Work done during internship at Microsoft Research Asia.}}
\address{$^1$Department of Computer Science and Technology, Tongji University\\
$^2$Department of Computer Science and Engineering, Southern University of Science and Technology\\
$^3$Department of Computing, The Hong Kong Polytechnic University\\
$^4$Microsoft Research Asia}
\begin{document}
\ninept
\maketitle
\begin{abstract}
Initially developed for natural language processing (NLP), Transformer model is now widely used for speech processing tasks such as speaker recognition, due to its powerful sequence modeling capabilities.
However, conventional self-attention mechanisms are originally designed for modeling textual sequence without considering the characteristics of speech and speaker modeling.
Besides, different Transformer variants for speaker recognition have not been well studied.
In this work, we propose a novel multi-view self-attention mechanism and present an empirical study of different Transformer variants with or without the proposed attention mechanism for speaker recognition.
Specifically, to balance the capabilities of capturing global dependencies and modeling the locality, we propose a multi-view self-attention mechanism for speaker Transformer, in which different attention heads can attend to different ranges of the receptive field.
Furthermore, we introduce and compare five Transformer variants with different network architectures, embedding locations, and pooling methods to learn speaker embeddings.
Experimental results on the VoxCeleb1 and VoxCeleb2 datasets show that the proposed multi-view self-attention mechanism achieves improvement in the performance of speaker recognition, and the proposed speaker Transformer network attains excellent results compared with state-of-the-art models.

\end{abstract}
\section{Introduction}
\label{sec:intro}


Transformer models \cite{transformer2017} have recently demonstrated exemplary performance on a broad range of natural language processing (NLP) tasks, such as machine translation 
and question answering.
Compared with recurrent neural networks (RNNs) and convolutional neural networks (CNNs), the advantage of self-attention in Transformer lies in its high parallelization capabilities and global modeling capabilities.
Recently, there have been increasing interests in exploring Transformers for spoken language processing, e.g., speech recognition \cite{Chen2021,zhou2021configurable,li2021recent}, speech synthesis \cite{Li_Liu_Liu_Zhao_Liu_2019,Huang2020}, and speaker recognition \cite{J2020,Safari2020}.



In speaker recognition, however, convolutional architectures remain dominant, such as residual network (ResNet) \cite{Cai2018,Xie2019,Chung2020} and time delay neural network (TDNN) \cite{Snyder2018,Okabel2018}. Inspired by the successes of self-attention in NLP, several works have tried to combine CNN-like architectures with self-attention by either replacing utterance-level pooling layers or frame-level convolutions blocks \cite{India2021,Zhu2021,Shi2020}.
Nevertheless, the overall structure of the previous work remains unchanged, and the application of Transformer to speaker recognition is limited. It is an interesting topic to explore effective ways of modeling speaker embedding with Transformer.

Applying Transformer to speaker tasks has two challenges: 1) Transformer is hard to be scaled efficiently since acoustic features sequences are much longer than text sentences. 2) Transformer is deficient in some of the inductive biases inherent to CNNs, such as translation equivalence and locality \cite{Dosovitskiy2020}. 
To enable Transformers to model the long-duration speech and locality, we propose a multi-view self-attention mechanism, where different attention heads can attend to different ranges of the receptive field to boost the capabilities of capturing global dependencies and modeling the locality.
Furthermore, we present a thorough empirical exploration of different Transformer models with different network architectures, embedding locations, and pooling methods for speaker recognition, equipped with our proposed multi-view self-attention mechanism.
We train the Transformer to represent speakers in a supervised speaker classification fashion, which encourages the encoder to capture different speaker properties by short-term or long-term dependencies. 
Experiments on the VoxCeleb1 and VoxCeleb2 datasets show that the proposed speaker Transformer network outperforms other CNNs and Transformer-based networks in that it achieves 96.38\% top-1 accuracy on the identification task and 2.56\% equal error rate on the verification task.

Our contributions can be summarized as follows. (1) We propose a multi-view self-attention mechanism for Transformer-based speaker networks, which enable to capture global dependencies and model the locality. (2) We study the proposed multi-view self-attention mechanism in different Transformer variants with different network structures, embedding locations, and pooling methods.

\section{Related Work}
\label{sec:related_work}

Transformers were proposed by Vaswani et al. \cite{transformer2017} for machine translation, and have become the state-of-the-art method in many NLP tasks. To apply Transformers in the context of speaker recognition, several works study this issue.

For speaker recognition, the attention mechanism has been studied with the pooling mechanism as an alternative to aggregate temporal information. 
Cai et al. \cite{Cai2018} introduce a self-attentive pooling layer to obtain the utterance-level representation. 
Okabe et al. \cite{Okabel2018} propose attentive statistics pooling, which gives different weights to different frames and generates weighted means and standard deviations. Wu et al. \cite{Wu2020} improve it by adopting a vectorial attention mechanism.
India et al. \cite{India2021} present double multi-head attention pooling, where an additional self-attention layer is added to the pooling layer to enhance the attentive pooling mechanism.
To improve the diversity of attention heads, Wang et al. \cite{Wang2020} propose multi-resolution multi-head attention pooling, which incorporates different resolutions of attentive weights.
Instead of using a fixed query for all utterances, Zhu et al. \cite{Zhu2021} introduce a self-attention mechanism with an input-aware query to consider overall information and speech dynamics over each utterance.

On the other hand, Jiang et al. \cite{Jiang2019} introduce a channel-wise attention mechanism as a gate, which can exploit global time-frequency information to improve the sensitivity of informative features while suppressing less useful ones.
Yu et al. \cite{Yu2020} propose a dynamic channel-wise selection mechanism based on the softmax attention to gather effective information and estimate the importance of network branches.
These works utilize the attention mechanism as a selection of channel-wise information in a block of the feature extractor. It is of limited worth for extracting speaker embedding.

Most recently, the attention layer has been directly stacked as a part of layers or the whole feature extractor.
On the top of the x-vectors framework \cite{Snyder2018}, Shi et al. \cite{Shi2020} apply Transformer encoders to both frame-level and segment-level to capture features at different scales.
Zhu et al. \cite{Zhu2021} propose a serialized multi-layer multi-head attention to obtain the final utterance-level embedding by aggregating the utterance-level vectors from all heads.
These works depend on the frame-level sophisticated convolutional networks such as TDNNBlocks \cite{Snyder2018} and SE-Res2Blocks \cite{Desplanques2020}. 
In contrast, Safari et al. \cite{Safari2020} propose a tandem self-attention encoding and pooling (SAEP) mechanism, which stacks two Transformer encoder layers followed by an additive attention pooling.
Metilda et al. \cite{J2020} propose s-vectors which stacks Transformer encoder layers followed by a statistics pooling layer and two linear layers.
However, Transformer-based feature extractors lack the capacity to model the locality and possess inferior performance in speaker recognition.


\section{Speaker Transformer}
\label{sec:speakertransformer}



\begin{figure}
 \centering
 \setlength{\abovecaptionskip}{3pt}
 \includegraphics[width=5.4cm]{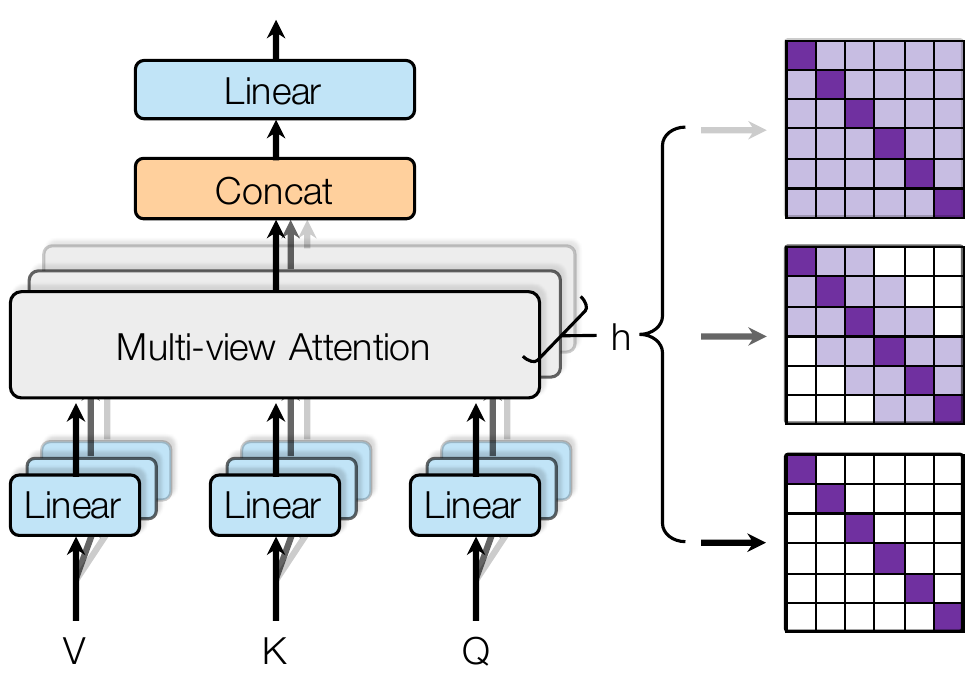}
 \caption{The proposed multi-view self-attention mechanism.}
 \label{fig:multi_view_attention}
\vspace{-12pt}\end{figure}

\subsection{Multi-View Self-Attention}

We propose a multi-view self-attention mechanism for Transformer to enhance the capabilities of capturing global dependencies while modeling the locality.
As shown in Fig \ref{fig:multi_view_attention}, the multi-view self-attention mechanism is implemented as self-attention with sliding windows of different sizes, in which each attention head has a different range of the receptive field.

Given the importance of local context, the proposed multi-view self-attention mechanism employs windows with different sizes surrounding each token. 
Using multiple stacked layers of such windowed attention creates various receptive fields, where top layers have access to long-range input locations.
Therefore, similar to CNNs, it can build representations that incorporate information across the input.
Specifically, given a fixed window size $w$, each token attends to $\frac{1}{2}w$ tokens on both sides. 
At the $l$-th layer of a Transformer encoder, the receptive field size ranges from $l\times w_\text{min}$ to $l \times w_\text{max}$, where $w_\text{min}$ and $w_\text{max}$ are the minimum and maximum window sizes for all layers, respectively.

For different Transformer variants, it might be helpful to use different values of $w_\text{min}$ and $w_\text{max}$ for each layer to model long-term or short-term dependencies.
However, it is computationally prohibitive to fine-tune the size of windows at each layer, because there is a vast search space of window size as the temporal length and layer number increase.
Intuitively, we simplify the selection of sliding window for the $i$-th head at the $l$-th layer to explicitly model different ranges of receptive fields by setting them as
\[
\setlength{\abovedisplayskip}{5pt}
\setlength{\belowdisplayskip}{5pt}
    w_i^l = 
\begin{cases}
    2^i+1,& \text{if } i\geq 1\\
    1,              & i = 0
\end{cases}.
\]

Given the matrices $Q$, $K$, and $V$ in the Transformer model \cite{transformer2017}, the proposed multi-view attention mechanism is formulated as
\begin{equation}
\setlength{\abovedisplayskip}{5pt}
\setlength{\belowdisplayskip}{5pt}
    \text{Attention}(Q, K, V) = M \odot \text{softmax}\left( \frac{QK^T}{\sqrt{d_k}} \right) V,
\end{equation}
where $M \in \mathbb{R}^{B \times H \times N \times N}$ is a head-wise masking matrix, $B$ is batch size, $H$ is the number of heads, $N$ is the number of steps, and $d_k$ is the dimension of queries and keys as mentioned in \cite{transformer2017}.


\subsection{Transformer Variants}


The general Transformer architecture used in machine translation \cite{transformer2017} consists of encoder blocks and decoder blocks. 
Each encoder block contains a multi-head attention and a feedforward network, while each decoder block has an additionally masked multi-head attention. 
All of the attention modules and feedforward networks are in conjunction with the residual connection and layer normalization.

We study five variants of the Transformer architecture for identifying speakers and extracting speaker embedding. We use an architecture with a 6-layer encoder, a 3-layer decoder, 512 attention size, 2048 hidden size, and 8 attention heads, which contain parameters up to 34.6 million. For all variants, the input $\mathbf{X}$ is firstly processed by two one-dimensional convolutional layers (called sub-sample encoder prenet) for downsampling to a quarter of the input length, followed by the Transformer encoder. 
Downsampling acoustic features $\mathbf{H}$ accelerates the processing efficiency of a speech utterance while forming coarse features to lay the base of extracting speaker-discriminative characteristics. In the following, we introduce the five variants.

\begin{figure*}
    \centering
    \setlength{\abovecaptionskip}{3pt}
    \subcaptionbox{First Decoder Token\label{fig:fivevariants_a}}{
        \includegraphics[height=2.8cm]{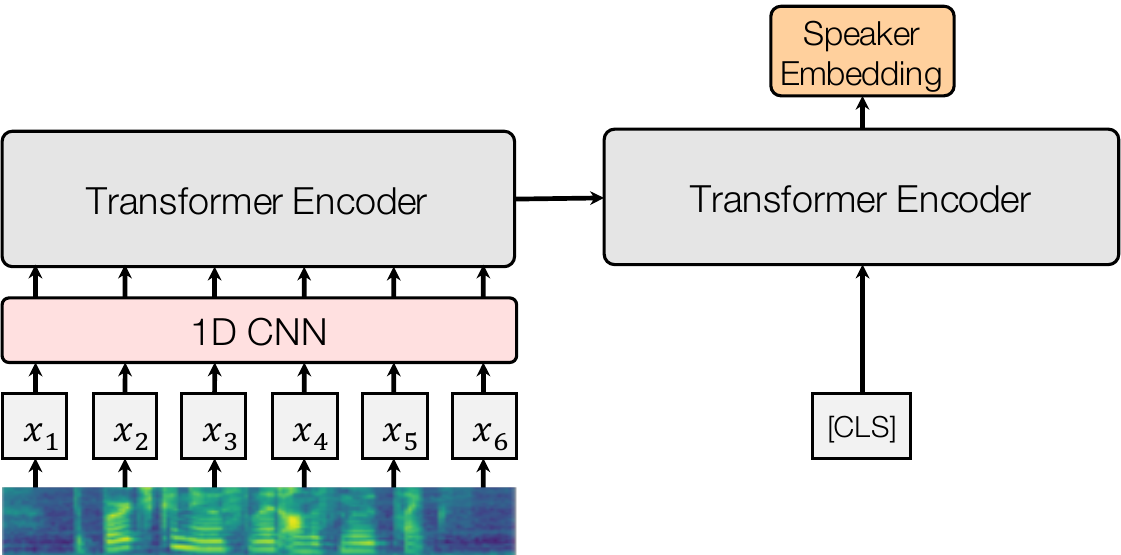}
    }\hspace{0.5cm}
    \subcaptionbox{Last Decoder Token\label{fig:fivevariants_b}}{
        \includegraphics[height=2.8cm]{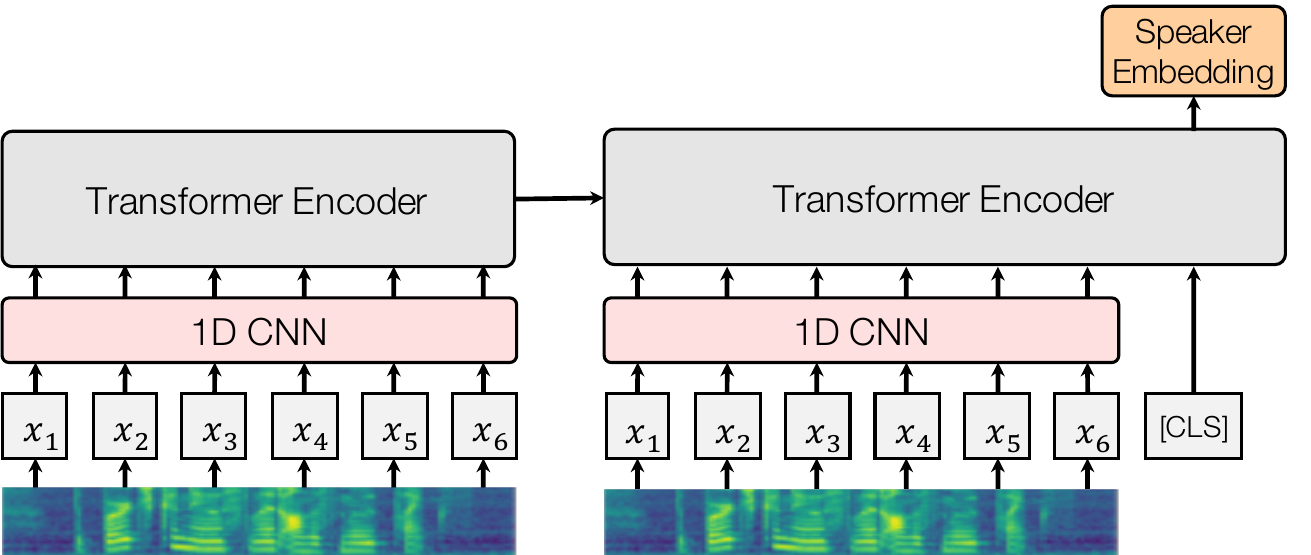}
    }\\
    \subcaptionbox{Average Encoder Tokens\label{fig:fivevariants_c}}{
        \includegraphics[width=3cm]{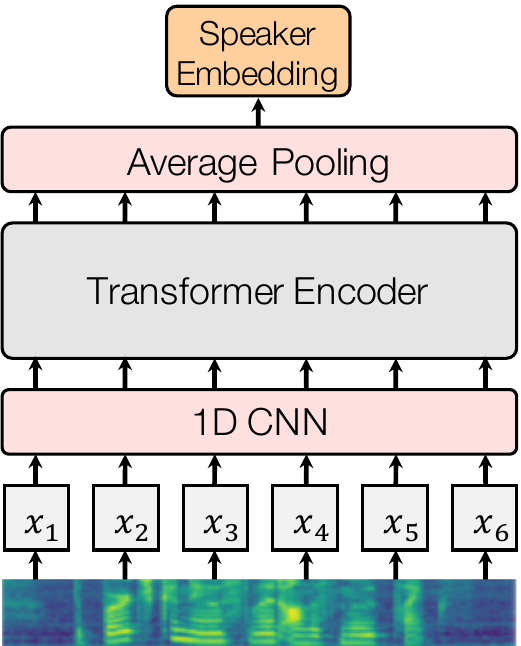}
    }\hspace{0.5cm}
    \subcaptionbox{First Encoder Token\label{fig:fivevariants_d}}{
        \includegraphics[width=4.15cm]{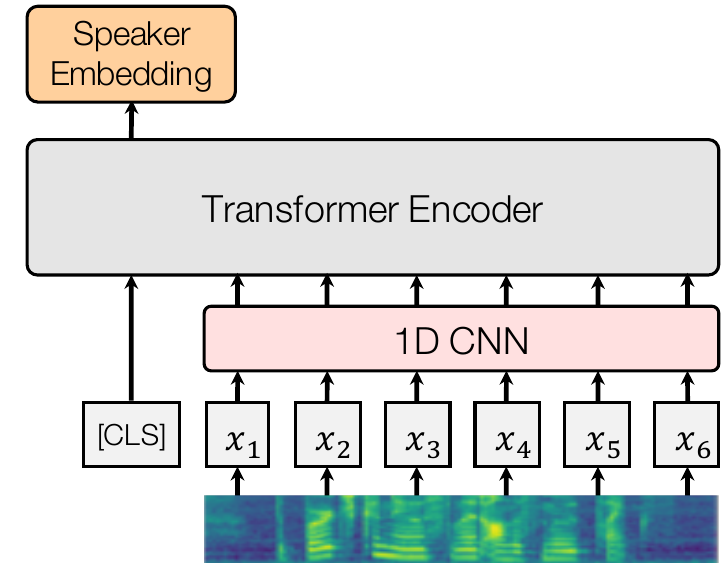}
    }\hspace{0.5cm}
    \subcaptionbox{Pooling Encoder Tokens\label{fig:fivevariants_e}}{
        \includegraphics[width=3cm]{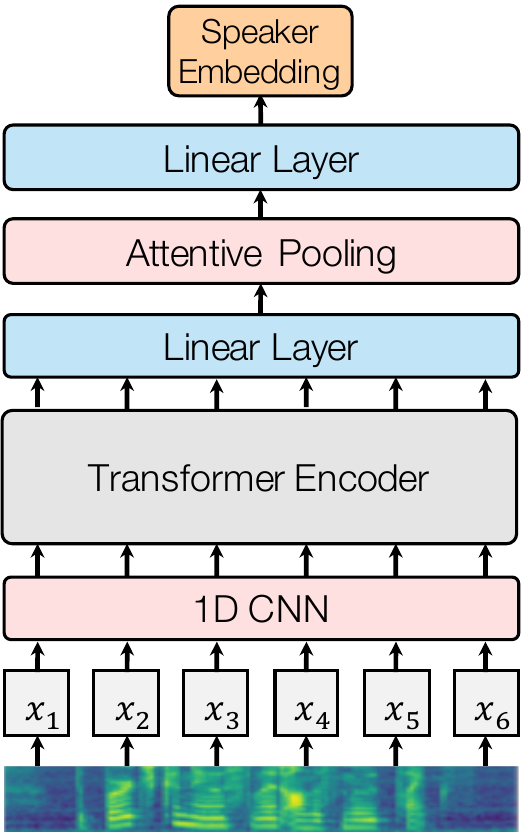}
    }
    \caption{Five Transformer variants for extracting speaker embedding.}
    \label{fig:fivevariants}
\vspace{-12pt}\end{figure*}
\begin{noindlist}
    \item[(a)]~\textbf{First Decoder Token}. As shown in Fig \ref{fig:fivevariants_a}, the Transformer architecture with an encoder and a decoder is considered as the speaker network.
    Specifically, the Transformer decoder and encoder take the \texttt{[CLS]} token and ${\mathbf{H}}$ as the input, respectively. 
    In this variant, the Transformer decoder acts as a multi-layer multi-head attentive pooling and takes advantage of stacked pooling, which is helpful to generate speaker-discriminative vectors.
    \item[(b)]~\textbf{Last Decoder Token}. Similar to \cite{lewis2020bart}, we formulate the problem of speaker classification as a sequence classification task.
    Different from the first variant, ${\mathbf{H}}$ is both fed into the encoder and decoder. The decoder can be considered as input-wise pooling.
    Then the final step of the decoder takes \texttt{[CLS]} token as input and generates speaker embedding, as shown in Fig \ref{fig:fivevariants_b}.
    \item[(c)]~\textbf{Average Encoder Token}. The naive temporal average pooling is directly applied without frame-level and utterance-level transformations to represent speakers. 
    The output of the Transformer encoder is averaged to obtain speaker embedding as shown in Fig \ref{fig:fivevariants_c}.
    \item[(d)]~\textbf{First Encoder Token}.
    As shown in Fig \ref{fig:fivevariants_d}, \texttt{[CLS]} token is concatenated with $\mathbf{H}$ as the input of the encoder, and the first output of the Transformer encoder is regarded as speaker embedding.
    Compared with other variants that utilize all tokens of the encoder, this variant utilizes a single token at the top layer of the encoder, which might cause the reducing diversity of temporal information.
    \item[(e)]~\textbf{Pooling Encoder Tokens}. Following the architectural setting in x-vector \cite{Snyder2018}, a linear layer, attentive pooling, and multiple hidden layers are sequentially stacked on the Transformer encoder, as shown in Fig \ref{fig:fivevariants_e}. 
    The difference between the x-vector and this Transformer variant is that the former uses a TDNN network as the feature extractor.
    Compared with \cite{J2020,Safari2020}, the linear transformation of inputs and temporal pooling are replaced by the sub-sample encoder prenet and attentive statistics pooling, respectively.
\end{noindlist}
\section{Experiments}
\label{sec:experiments}

\subsection{Setup}
\label{ssec:setup}

\textbf{Dataset.} We focus on text-independent speaker recognition and use the VoxCeleb dataset to evaluate the performance on both speaker identification and verification tasks. 
The VoxCeleb dataset, containing VoxCeleb1 \cite{Nagrani2017} and VoxCeleb2 \cite{Chung2018}, is a large-scale text-independent speaker recognition dataset collected ``in the wild". 
The VoxCeleb1 has over 100,000 utterances from 1,251 celebrities, while the VoxCeleb2 has over 1,000,000 utterances from 6,112 celebrities.
We use the official split of VoxCeleb1 for the speaker identification task, where the test set contains 8,251 utterances from these 1,251 celebrities.
For the speaker verification task, we consider two settings. 
The VoxCeleb1 with 1,211 speakers and VoxCeleb2 with 5,994 speakers are used for training, respectively. 
The test set contains 4,715 utterances from 40 speakers in VoxCeleb1. 
There are 37,720 pairs of trials, including 18,860 target pairs.

\hspace{-14.6pt}\textbf{Acoustic Features.} In our experiments, the \emph{librosa} toolkit is used to extract 80-dimensional mel-filter banks with the 64ms window and 16ms shift to represent the speech signal. 
No data augmentation is used during the training or test process. All features are subject to 200-frame utterance-level cepstral mean variance normalization.
We apply SpecAugment \cite{Park2019} which randomly masks 0 to 20 frames up to twice in the time domain and 0 to 10 frequency banks up to twice.



\hspace{-14.6pt}\textbf{Training and Metrics.} Using \emph{fairseq} \cite{ott2019fairseq}, we train all models on both tasks via the Adam optimizer with $\beta_1=0.9$, $\beta_2=0.999$, a weight decay of 0.1, a batch of 2048, the inputs of 200 frames, and the dropout of 0.1. 
We adjust the learning rate based on a 60k-step cycle of a triangular cyclical schedule between $10^{-8}$ and $5\times10^{-4}$. 
Four cycles are applied to VoxCeleb2 while 2 cycles is to VoxCeleb1 due to a smaller data size.
During the test stage, speaker embeddings are extracted from the whole speech signal. 
We report the top-1 accuracy (ACC) as the identification metric and the equal error rate (EER) for verification.

\subsection{Different Transformer Variants}

The results of five variants with or without multi-view self-attention (MV)
are shown in Table \ref{tab:transformer_variants}, where ACC on VoxCeleb1 and EER on VoxCeleb1 and VoxCeleb2 are reported. 
Except variant (d), most Transformer variants with MV outperform those without MV, which indicates the proposed MV is helpful to improve the performance.
Specifically, training on middle-scale corpus VoxCeleb1, MV achieves improvements in most tasks for variants (a), (b), (c), and (e) but slightly degrades in the verification task for variant (a) and identification task for variant (c).
As the size of the dataset scales, the performance of speaker embedding boosts consistently and attains 5.9\%-10.3\% improvement for variants (a), (b), (c), and (e). 
It suggests that MV is an effective self-attention enhancement technique that can be jointly used with other techniques such as attentive statistics pooling in variant (e).

On the other hand, the performance of variant (d) with MV significantly degrades regardless of the scale of datasets varies. 
This variant uses the first token at the top layer of the encoder as speaker embedding. 
The self-attention with different sliding window masks forces heads on the importance of various local context, which makes heads learn different distributions of temporal information and contribute unequally to extract speaker embedding. 
It causes that variant (d) suffers the loss of part of the temporal information.

\begin{table}
 \centering
 \setlength{\abovecaptionskip}{3pt}
 \setlength{\belowcaptionskip}{-0pt}
 \caption{Performance of MV-based Transformer on VoxCeleb1.}
 \begin{tabular}{crrrrrr}
 \toprule
 Variants & \multicolumn{2}{c}{ACC (\%) $\uparrow$} & \multicolumn{2}{c}{EER (\%) $\downarrow$} & \multicolumn{2}{c}{EER (\%) $\downarrow$} \\
 \cmidrule(r){2-3}\cmidrule(r){4-5}\cmidrule(r){6-7}
 & Vox1 & + MV & Vox1 & + MV & Vox2 & + MV \\
 \midrule
 (a) & 94.33 & \textbf{94.36} & \textbf{5.33} & 5.45 & 2.72 & \textbf{2.56} \\
 (b) & 93.61 & \textbf{94.09} & 5.89 & \textbf{5.40} & 2.92 & \textbf{2.68} \\
 (c) & \textbf{92.96} & 91.81 & 6.33 & \textbf{6.13} & 3.60 & \textbf{3.23} \\
 (d) & \textbf{92.29} & 88.16 & \textbf{5.96} & 7.37 & \textbf{3.32} & 3.96 \\
 (e) & 95.04 & \textbf{96.38} & 4.77 & \textbf{4.35} & 2.89 & \textbf{2.68} \\
 \bottomrule
 \end{tabular}
 \label{tab:transformer_variants}
\vspace{-12pt}\end{table}

Except variant (d), the Transformer variants with or without MV give the same ranking of the verification performance on the VoxCeleb2, i.e., (a), (e), (b), (c). 
Variant (c) introduces a naive temporal average pooling among features derived from the Transformer encoder, which represents the fundamental  capability of identifying the speaker.
On the top of the transformer encoder, variant (a) applies multi-head attentive pooling, variant (e) applies attentive statistics pooling, and variant (c) applies input-wise pooling to achieve superior performance. 
Although variant (b) provides an additional input, it is inferior to variant (a).
It is probably that the additional input causes over-regularization to the decoder, whose weights require learning the mapping function from both the input and \texttt{[CLS]} token to the speaker identity. 
By considering that the temporal standard in the utterance-level features is complementary to the temporal average pooling, it implies that combining multi-head attentive pooling and statistics pooling can further improve the performance.

\subsection{Comparison on VoxCeleb}
\label{ssec:comparison}

We compare the proposed method with several models, including VGG \cite{Nagrani2017}, TDNN \cite{Snyder2018,Okabel2018}, ResNet \cite{Cai2018,Xie2019,Chung2020}, and Transformer \cite{J2020,Safari2020}. According to results on the VoxCeleb speaker recognition tasks shown in Table \ref{tab:comparison_vox}, we can see that the proposed method outperforms most works using convolutional network or attention mechanism on three speaker tasks. 
Compared with those methods based on VGG, TDNN, and ResNet as the feature extractor, the proposed Transformer encoder stacked on a sub-sample prenet attains excellent performance in both ACC and EER.
To the best of our knowledge, the obtained ACC is the state-of-the-art performance, which indicates that the Transformer-based speaker network possesses superior capability for classification. 
On the other hand, our work is inferior to the TDNN with attentive statistics pooling \cite{Okabel2018}. 
It is probably that the data augmentation technique increases the diversity of training dataset, which is helpful to generalize to unseen speakers and unseen acoustic scenes. 

Regarding the feature extractor, variant (a) outperforms the VGG with multi-head attention \cite{India2021}, and variant (e) achieves comparable performance with the TDNN equipped with attentive statistics pooling \cite{Okabel2018}. 
It suggests that compared with several popular CNNs, the Transformer encoder has a comparative capability to extract frame-level features for generating speaker-discriminative embeddings.

\begin{table}
 \centering
 \setlength{\abovecaptionskip}{3pt}
 \setlength{\belowcaptionskip}{-0pt}
 \caption{Performance comparison on the VoxCeleb1 test Set.}
 \begin{tabular}{lrrr}
 \toprule
 \multicolumn{4}{c}{\textbf{Training on VoxCeleb1 development}} \\
 Implementaion & Extractor & ACC (\%) & EER (\%) \\
 \midrule
 VGG-M \cite{Nagrani2017} & VGG & 80.5 & 7.8 \\
 X-vector \cite{Snyder2018} & TDNN & - & 7.83 \\
 Atten. Stats.*\cite{Okabel2018} & TDNN & - & \textbf{3.85} \\
 Cai et al. \cite{Cai2018} & ResNet & 89.9 & 4.46 \\
 Chung et al. \cite{Chung2020} & ResNet & 89.0 & 5.26 \\
 SAEP \cite{Safari2020} & Transformer & - & 7.13 \\
 S-vectors \cite{J2020} & Transformer & - & 5.50 \\
 \midrule
 Our work (e) & CNN+Transformer & \textbf{96.38} & 4.35 \\
\midrule
\midrule
\multicolumn{4}{c}{\textbf{Training on VoxCeleb2 development}} \\
 MHA \cite{India2021} & VGG & & 3.19 \\
 Atten. Stats. \cite{Okabel2018} & TDNN & & 2.59 \cite{Wu2020} \\
 Xie et al. \cite{Xie2019} & ResNet & & 3.22 \\
 
 SAEP \cite{Safari2020} & Transformer & & 5.44 \\
 S-vectors*$^{+}$\cite{J2020} & Transformer & & 2.67 \\
 \midrule
 Our work (a) & CNN+Transformer & & \textbf{2.56} \\
 Our work (e) & CNN+Transformer & & 2.68 \\
 \bottomrule
 \multicolumn{4}{l}{* Training using data augmentation.} \\
 \multicolumn{4}{l}{$^{+}$ Training dataset includes VoxCeleb2 and VoxCeleb1 dev set.} \\
 \end{tabular}
 \label{tab:comparison_vox}
\vspace{-12pt}\end{table}

For the Transformer-based speaker verification, the proposed method achieves EER of 4.35\% and 2.56\% when training on the VoxCeleb1 and VoxCeleb2, respectively. 
We further boost the performance based on \cite{J2020,Safari2020} where the Transformer encoder is applied to extract features. 
For example, considering that SEAP \cite{J2020} designs a lightweight network with 1.60 million parameters, one reason for the improvement of the proposed method is the model scaling. 
Regardless of the small size, it often does not scale effectively as the length of inputs increases.
Therefore, the sub-sample prenet is employed in our work, which leads to a significant reduction in terms of storage size, processing, and memory.
S-vectors \cite{Safari2020} trained on the VoxCeleb1 and VoxCeleb2 development sets and data augmentation is inferior to the proposed method. 
It suggests several architectural enhancements to the Transformer-based speaker network such as attentive pooling, multi-head pooling, and multi-layer pooling.

\section{Conclusion}
\label{sec:conclusion}


In this work, we explore five Transformer variants for speaker recognition. A multi-view self-attention mechanism is proposed to balance the capabilities of capturing global dependencies and modeling the locality by using sliding windows with different sizes for each attention head.
The proposed attention mechanism achieves improvements on most variants for both speaker identification and speaker verification tasks.
Moreover, the proposed model attains excellent results compared to several previous CNN-based and Transformer-based models.
Our method achieves 96.38\% top-1 accuracy for the speaker identification task on Voxceleb1, which is state-of-the-art to the best of our knowledge, and 4.35\% and 2.56\% EER on VoxCeleb1 and VoxCeleb2, respectively, for the speaker verification task.
In the future work, we will utilize larger datasets by pretraining techniques
\cite{Hsu2021,ao2021speecht5}
and employ data augmentation techniques \cite{tom2015speed,ko2017study} to further boost the performance.

\section{Acknowledgements}
\label{sec:acknowledgements}

This work is partially supported by the National Nature Science Foundation of China (No. 61976160, 62076182, 61906137) and Technology research plan project of Ministry of Public and Security (Grant No. 2020JSYJD01) and Shanghai Science and Technology Plan Project (No. 21DZ1204800).





\bibliographystyle{IEEEbib}
\bibliography{strings,refs}

\begin{thebibliography}{10}

\bibitem{transformer2017}
Ashish Vaswani, Noam Shazeer, Niki Parmar, Jakob Uszkoreit, Llion Jones,
  Aidan~N Gomez, \L~ukasz Kaiser, and Illia Polosukhin,
\newblock ``Attention is all you need,''
\newblock in {\em Proc. NeurIPS}, 2017, pp. 6000--6010.

\bibitem{Chen2021}
Xie Chen, Yu~Wu, Zhenghao Wang, Shujie Liu, and Jinyu Li,
\newblock ``{Developing real-time streaming transformer transducer for speech
  recognition on large-scale dataset},''
\newblock in {\em Proc. ICASSP}, 2021, pp. 5904--5908.

\bibitem{zhou2021configurable}
Long Zhou, Jinyu Li, Eric Sun, and Shujie Liu,
\newblock ``A configurable multilingual model is all you need to recognize all
  languages,''
\newblock {\em arXiv preprint arXiv:2107.05876}, 2021.

\bibitem{li2021recent}
Jinyu Li,
\newblock ``Recent advances in end-to-end automatic speech recognition,''
\newblock {\em arXiv preprint arXiv:2111.01690}, 2021.

\bibitem{Li_Liu_Liu_Zhao_Liu_2019}
Naihan Li, Shujie Liu, Yanqing Liu, Sheng Zhao, and Ming Liu,
\newblock ``{Neural speech synthesis with transformer network},''
\newblock in {\em Proc. AAAI}, 2019, pp. 6706--6713.

\bibitem{Huang2020}
Wen-Chin Huang, Tomoki Hayashi, Yi-Chiao Wu, Hirokazu Kameoka, and Tomoki Toda,
\newblock ``{Voice Transformer Network: Sequence-to-sequence voice conversion
  using transformer with text-to-speech pretraining},''
\newblock in {\em Proc. Interspeech}, 2020, pp. 4676--4680.

\bibitem{J2020}
Metilda Sagaya Mary~N J, Sandesh~V Katta, and S~Umesh,
\newblock ``{S-vectors: Speaker embeddings based on transformer's encoder for
  text-independent speaker verification},''
\newblock {\em arXiv preprint arXiv:2008.04659}, 2020.

\bibitem{Safari2020}
Pooyan Safari, Miquel India, and Javier Hernando,
\newblock ``Self-attention encoding and pooling for speaker recognition,''
\newblock in {\em Proc. Interspeech}, 2020, pp. 941--945.

\bibitem{Cai2018}
Weicheng Cai, Jinkun Chen, and Ming Li,
\newblock ``{Exploring the encoding layer and loss function in end-to-end
  speaker and language recognition system},''
\newblock in {\em Proc. Odyssey}, 2018, pp. 74--81.

\bibitem{Xie2019}
Weidi Xie, Arsha Nagrani, Joon~Son Chung, and Andrew Zisserman,
\newblock ``{Utterance-level aggregation for speaker recognition in the
  wild},''
\newblock in {\em Proc. ICASSP}, 2019, pp. 5791--5795.

\bibitem{Chung2020}
Joon~Son Chung, Jaesung Huh, and Seongkyu Mun,
\newblock ``{Delving into VoxCeleb: Environment invariant speaker
  recognition},''
\newblock in {\em Proc. Odyssey}, 2020, pp. 349--356.

\bibitem{Snyder2018}
David Snyder, Daniel Garcia-Romero, Gregory Sell, Daniel Povey, and Sanjeev
  Khudanpur,
\newblock ``{X-vectors: Robust DNN embeddings for speaker recognition},''
\newblock in {\em Proc. ICASSP}, 2018, pp. 5329--5333.

\bibitem{Okabel2018}
Koji Okabe, Takafumi Koshinaka, and Koichi Shinoda,
\newblock ``{Attentive statistics pooling for deep speaker embedding},''
\newblock in {\em Proc. Interspeech}, 2018, pp. 2252--2256.

\bibitem{India2021}
Miquel India, Pooyan Safari, and Javier Hernando,
\newblock ``{Double multi-head attention for speaker verification},''
\newblock in {\em Proc. ICASSP}, 2021, pp. 6144--6148.

\bibitem{Zhu2021}
Hongning Zhu, Kong~Aik Lee, and Haizhou Li,
\newblock ``{Serialized multi-layer multi-head attention for neural speaker
  embedding},''
\newblock {\em arXiv preprint arXiv:2107.06493}, 2021.

\bibitem{Shi2020}
Yanpei Shi, Mingjie Chen, Qiang Huang, and Thomas Hain,
\newblock ``T-vectors: Weakly supervised speaker identification using
  hierarchical transformer model,''
\newblock {\em arXiv preprint arXiv:2010.16071}, 2020.

\bibitem{Dosovitskiy2020}
Alexey Dosovitskiy, Lucas Beyer, Alexander Kolesnikov, Dirk Weissenborn,
  Xiaohua Zhai, Thomas Unterthiner, Mostafa Dehghani, Matthias Minderer, Georg
  Heigold, Sylvain Gelly, Jakob Uszkoreit, and Neil Houlsby,
\newblock ``{An image is worth 16x16 words: Transformers for image recognition
  at scale},''
\newblock {\em arXiv preprint arXiv:2010.11929}, 2020.

\bibitem{Wu2020}
Yanfeng Wu, Chenkai Guo, Hongcan Gao, Xiaolei Hou, and Jing Xu,
\newblock ``{Vector-based attentive pooling for text-independent speaker
  verification},''
\newblock in {\em Proc. Interspeech}, 2020, pp. 936--940.

\bibitem{Wang2020}
Zhiming Wang, Kaisheng Yao, Xiaolong Li, and Shuo Fang,
\newblock ``{Multi-resolution multi-head attention in deep speaker
  embedding},''
\newblock in {\em Proc. ICASSP}, 2020, pp. 6464--6468.

\bibitem{Jiang2019}
Yiheng Jiang, Yan Song, Ian McLoughlin, Zhifu Gao, and Lirong Dai,
\newblock ``{An effective deep embedding learning architecture for speaker
  verification},''
\newblock in {\em Proc. Interspeech}, 2019, pp. 4040--4044.

\bibitem{Yu2020}
Ya-Qi Yu and Wu-Jun Li,
\newblock ``{Densely connected time delay neural network for speaker
  verification},''
\newblock in {\em Proc. Interspeech}, 2020, pp. 921--925.

\bibitem{Desplanques2020}
Brecht Desplanques, Jenthe Thienpondt, and Kris Demuynck,
\newblock ``{ECAPA-TDNN: Emphasized channel attention, propagation and
  aggregation in TDNN based speaker verification},''
\newblock in {\em Proc. Interspeech}, 2020, pp. 3830--3834.

\bibitem{lewis2020bart}
Mike Lewis, Yinhan Liu, Naman Goyal, Marjan Ghazvininejad, Abdelrahman Mohamed,
  Omer Levy, Veselin Stoyanov, and Luke Zettlemoyer,
\newblock ``{BART}: Denoising sequence-to-sequence pre-training for natural
  language generation, translation, and comprehension,''
\newblock in {\em Proc. ACL}, 2020, pp. 7871--7880.

\bibitem{Nagrani2017}
Arsha Nagrani, Joon~Son Chung, and Andrew Zisserman,
\newblock ``{VoxCeleb: A large-scale speaker identification dataset},''
\newblock in {\em Proc. Interspeech}, 2017, pp. 2616--2620.

\bibitem{Chung2018}
Joon~Son Chung, Arsha Nagrani, Andrew Zisserman, and Assoc Int~Speech Commun,
\newblock ``Voxceleb2: Deep speaker recognition,''
\newblock in {\em Proc. Interspeech}, 2018, pp. 1086--1090.

\bibitem{Park2019}
Daniel~S. Park, William Chan, Yu~Zhang, Chung~Cheng Chiu, Barret Zoph, Ekin~D.
  Cubuk, and Quoc~V. Le,
\newblock ``{Specaugment: A simple data augmentation method for automatic
  speech recognition},''
\newblock in {\em Proc. Interspeech}, 2019, pp. 2613--2617.

\bibitem{ott2019fairseq}
Myle Ott, Sergey Edunov, Alexei Baevski, Angela Fan, Sam Gross, Nathan Ng,
  David Grangier, and Michael Auli,
\newblock ``fairseq: A fast, extensible toolkit for sequence modeling,''
\newblock in {\em Proc. NAACL-HLT}, 2019.

\bibitem{Hsu2021}
Wei-Ning Hsu, Benjamin Bolte, Yao-Hung~Hubert Tsai, Kushal Lakhotia, Ruslan
  Salakhutdinov, and Abdelrahman Mohamed,
\newblock ``{HuBERT: Self-supervised speech representation learning by masked
  prediction of hidden units},''
\newblock {\em arXiv preprint arXiv:2106.07447}, 2021.

\bibitem{ao2021speecht5}
Junyi Ao, Rui Wang, Long Zhou, Shujie Liu, Shuo Ren, Yu~Wu, Tom Ko, Qing Li,
  Yu~Zhang, Zhihua Wei, Yao Qiao, Jinyu Li, and Furu Wei,
\newblock ``Speecht5: Unified-modal encoder-decoder pre-training for spoken
  language processing,''
\newblock {\em arXiv preprint arXiv:2110.07205}, 2021.

\bibitem{tom2015speed}
Tom Ko, Vijayaditya Peddinti, Daniel Povey, and Sanjeev Khudanpur,
\newblock ``Audio augmentation for speech recognition,''
\newblock in {\em Proc. Interspeech}, 2015, pp. 3586--3589.

\bibitem{ko2017study}
Tom Ko, Vijayaditya Peddinti, Daniel Povey, Michael~L Seltzer, and Sanjeev
  Khudanpur,
\newblock ``A study on data augmentation of reverberant speech for robust
  speech recognition,''
\newblock in {\em Proc. ICASSP}, 2017, pp. 5220--5224.

\end{thebibliography}

\end{document}